\documentclass[twocolumn,twoside,aps,groupaddress,pre,showpacs,floatfix]{revtex4}

\usepackage{amsmath,amssymb,graphicx,latexsym}
\usepackage[dvips]{color}

\newcommand{\Ha}{{\mathbf H}}
\newcommand{\M}{{\mathbf M}}
\newcommand{\B}{{\mathbf B}}

\newcommand{\cH}{{\cal H}}

\newcommand{\phiu}{\Phi^{(1)}}
\newcommand{\phim}{\Phi^{(2)}}
\newcommand{\phio}{\Phi^{(3)}}
\newcommand{\rhou}{\rho^{(1)}}
\newcommand{\rhom}{\rho^{(2)}}
\newcommand{\rhoo}{\rho^{(3)}}
\newcommand{\sigmau}{\sigma^{(0)}}
\newcommand{\sigmao}{\sigma^{(d)}}
\newcommand{\zetau}{\zeta^{(0)}}
\newcommand{\zetao}{\zeta^{(d)}}
\newcommand{\E}{\mathrm{e}}
\newcommand{\D}{\mathrm{d}}
\renewcommand{\vec}[1]{\mathbf{#1}}

\begin{document}

\title{Double Rosensweig instability in a ferrofluid sandwich structure}

\author{Dirk Rannacher}
\email[]{rannacher@theorie.physik.uni-oldenburg.de}
\affiliation{Universit\"at Magdeburg, Institut f\"ur Theoretische
Physik,
  PSF 4120, 39106 Magdeburg, Germany}
\author{Andreas Engel}
\affiliation{Universit\"at Magdeburg, Institut f\"ur Theoretische
Physik,
PSF 4120, 39106 Magdeburg, Germany}

\date{\today}

\begin{abstract}
We consider a horizontal ferrofluid layer sandwiched between two layers of
immiscible non-magnetic fluids. In a sufficiently strong vertical 
magnetic field the flat interfaces between magnetic and non-magnetic
fluids become unstable to the formation of peaks. We theoretically 
investigate the interplay between these two instabilities for
different combinations of the parameters of the fluids and analyze the
evolving interfacial patterns. We also estimate the critical magnetic
field strength at which thin layers disintegrate into an ordered array
of individual drops.
\end{abstract}

\pacs{75.50.Mm,47.20.Ma}

\maketitle

\section{Introduction}
Ferrofluids are colloidal suspensions of nano-size ferromagnetic grains
in a carrier liquid like water or oil \cite{Ros}. The dipole-dipole
interaction between the ferromagnetic particles is for moderate volume
concentrations rather small and ferrofluids hence behave magnetically
as superparamagnets.  Accordingly in the absence of an external
magnetic field the magnetization of the fluid is zero. If a field is
switched on the magnetic moments of the particles orient themselves
along the field direction giving rise to a macroscopic
magnetization. The notion {\it super}paramagnets refers to the
unusually high value of the magnetic susceptibility, 
$\chi=1....50$, to be compared with $\chi\simeq 10^{-4}$ for atomic
paramagnets. Hydrodynamically diluted ferrofluids behave
like ordinary Newtonian liquids with additional contributions in the
bulk and surface force densities stemming from the interaction with the
magnetic field. 

Due to the unique interplay between hydrodynamic and magnetic degrees
of freedom ferrofluids show a variety of instabilities and pattern
formation processes. Among the most striking phenomena in this respect
is the so-called Rosensweig instability in which the flat free
surface of a ferrofluid becomes unstable when subjected to a
sufficiently strong vertical magnetic field \cite{CoRo}. Although 
both gravity and surface tension favor a flat surface the decrease in
magnetic energy for a periodic array of peaks and troughs can be large
enough to overcompensate the increase in potential and surface energy. Both
the linear instability and the details of the pattern formation as 
revealed by a weakly non-linear analysis have been thoroughly
studied \cite{CoRo,ZaSh,EnLaCh,Gal,KuSp,FrEn}.

In the present paper we investigate a sandwich structure
in which a ferrofluid layer of given thickness is placed between 
two immiscible non-magnetic liquids. The system is prepared such that
in the absence of a magnetic field the layering is stable, i.e. 
the lower layers have larger densities than the upper ones
in order to prevent the Rayleigh-Taylor instability. Applying a
homogeneous external magnetic field perpendicular to the undisturbed
interfaces gives rise to Rosensweig
instabilities at {\em both the lower and the upper} interface of the
ferrofluid layer. Due to the non-local character of the magnetic field
energy these instabilities are {\em coupled} with each other. We first 
study the interplay and competition between these instabilities within
the framework of the linear stability
analysis. Depending on the parameters of the system one interface
dominates and ``slaves'' the other one to its unstable wavenumber or
both interfaces become unstable at rather similar values of the
magnetic field giving rise to a competition between the corresponding
wavenumbers. This is similar to what occurs in
Rayleigh-B\'enard-Marangoni convection in systems of two superimposed
fluids which are coupled viscously and thermally at their common
interface \cite{RaBuRe,vanHook,ToMoMo,EnSw}.  

In order to characterize the patterns evolving from the instability we
perturbatively probe into the weakly non-linear regime by expanding
the free energy of the system in powers of the amplitude of the
surface deflections generalizing the methods developed in 
\cite{Gal,FrEn,Rene}. When the amplitude of the surface deformations
becomes comparable to the thickness of the ferrofluid layer itself the
layer may be decomposed into disconnected parts. Within our non-linear
analysis we are able to estimate the field strength necessary for such
a disintegration to occur. Finally by using 
experimentally relevant values for the parameters we point out
interesting experimental realizations of our system. 

The main difference between our sandwich system and the somewhat related
problem of a ferrofluid {\em film} investigated in
\cite{BaPeSa} 
is the thickness of the ferrofluid layer. For a film
this thickness is 
by definition much smaller than the wavelength of the unstable
mode. In the experiments reported in \cite{BaPeSa} the film thickness
varied between 5 and 60 $\mu$m. The hydrodynamics of the film can then
be very well described within the lubrication approximation. In our
system the thickness of the ferrofluid layer is comparable to the
unstable wavelength which is of the order of centimeters and
correspondingly the full hydrodynamic equations have to be solved to
describe its dynamics.  

\section{Basic equations}

\begin{figure}[t]
  \includegraphics[scale=1]{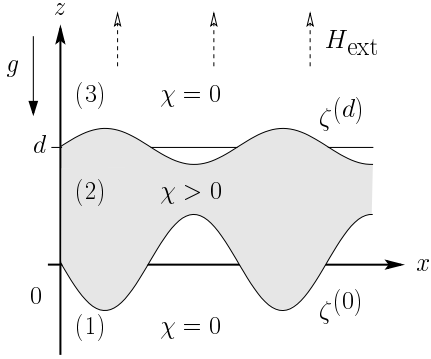}
  \caption{Schematic two dimensional plot of a ferrofluid layer of depth 
    $d$ with infinite horizontal extension sandwiched 
    between two non-magnetic liquids of infinite depth. }
  \label{rosensweig}
\end{figure}

We consider a horizontally unbounded ferrofluid layer of thickness $d$ 
and density $\rhom$ sandwiched between two immiscible, non-magnetic 
liquids with densities $\rhou$ and $\rhoo$. The interfaces 
between the layers are parametrized by the functions $z=\zetao(x)$ and 
$z=\zetau(x)$ where for simplicity we will only consider 
one-dimensional interface modulations (see fig.\ref{rosensweig}). It has 
recently been clarified that this situation can be realized 
experimentally by using an {\em oblique} magnetic field \cite{Rene2}. 
The interface tensions at the two interfaces are denoted by 
$\sigmao$ and $\sigmau$. 

The hydrodynamics of the system is quite generally described by the 
Navier-Stokes equation and the continuity equation. However, since the 
situation of interest is a static one these equations can be replaced by 
the pressure equilibrium at the two interfaces. This in turn is 
equivalent to the minimum condition for the total energy functional.

The energy per area in the $x$-$y$ plane comprises three parts, $E_h,\
E_s$, and $E_m$ denoting the 
hydrostatic, the interfacial, and the magnetic energies
respectively. The first two parts are given by the 
well-known expressions 
\begin{equation}
 E_h = g\left\langle\rhou\!\!\!\!
             \int\limits_{-\infty}^{\zetau(x)} \!\!\!\!\D z z
    + \rhom\!\!\!\!\!\int\limits_{\zetau(x)}^{\zetao(x)} \!\!\!\!\D z z 
    + \rhoo\!\!\!\!\!\int\limits_{\zetao(x)}^{\infty}\!\!\!\! \D z z
       \right\rangle
\label{E_h}
\end{equation}
and 
\begin{equation}
\begin{split}
 E_s &=\left\langle \sigma^{(0)}\sqrt{1+(\partial_x\zeta^{(0)}(x))^2}
      + \sigma^{(d)}\sqrt{1+(\partial_x\zeta^{(d)}(x))^2}\right\rangle
\end{split}
\label{E_s}
\end{equation}
where $g$ is the acceleration due to gravity. The brackets $\langle
... \rangle$ denote the spatial average along the $x$-direction
\begin{equation}
  \left\langle\dots\right\rangle=\lim_{L\rightarrow\infty} 
       \frac{1}{2L}\int_{-L}^{L}\D x\,\cdots\qquad.
  \label{mittelwert}
\end{equation}
The volume density of magnetic energy is of the general form \cite{LL} 
\begin{equation}
  e_m=-\mu_0 \int_0^{\vec{H_0}} \D \vec{H'}\cdot
  \vec{M}(\vec{H'}) 
\end{equation}
where $\vec{M}$ denotes the magnetization, $\vec{H_0}$ the magnetic
field {\it in the absence} of any permeable material, and $\mu_0$ is the
permeability of free space. Assuming a linear magnetization law
$\M=\chi \Ha$ of the ferrofluid with the susceptibility $\chi$
characterizing its magnetic properties we hence find in the present
case for the magnetic energy per unit area 
\begin{equation}\label{E_m}
 E_m=-\frac{\mu_0\;
   \chi}{2}\left\langle\int\limits_{\zetau(x)}^{\zetao(x)}
     \D z\;\vec{H}(x,z)\cdot\vec{H}_{\mathrm{ext}}\right\rangle.
\end{equation}
Here $\Ha_\mathrm{ext}$ denotes the homogeneous external magnetic
field produced by the experimental setup and in the absence of the
ferrofluid and $\Ha(x,z)$ is the actual magnetic field in the
ferrofluid.  

Subtracting an irrelevant constant the complete energy functional of
the system can hence be written as 
\begin{widetext}
\begin{multline}
    E\!\left[\zetau(x),\zetao(x)\right]=\\
    \left\langle\frac{g}{2}\left[(\rho^{(1)}-\rho^{(2)})
        \zeta^{(0)^2} + \left(\rho^{(2)}-\rho^{(3)}\right)
        \zeta^{(d)^2}\right]
 -\frac{\mu_0\;\chi}{2}\!\!\!\!\!\int
     \limits_{\zetau(x)}^{\zetao(x)}\!\!\!\!\!\D z\;  
      \vec{H}(x,z)\cdot\vec{H}_{\mathrm{ext}}
    + \sigma^{(0)}\sqrt{1+(\partial_x\zeta^{(0)}(x))^2}
      + \sigma^{(d)}\sqrt{1+(\partial_x\zeta^{(d)}(x))^2}
    \right\rangle
  \label{energief}
\end{multline}
\end{widetext}

The magnetic field has to obey the magnetostatic Maxwell equations 
\begin{equation}
 \nabla\cdot\vec{B}=0\quad\mbox{and}\quad\nabla\times\vec{H}=0
 \label{maxwell}
\end{equation}
with $\B=\mu_0(1+\chi) \Ha$. These equations are completed by the 
following boundary conditions
\begin{equation}
  \lim_{z\rightarrow\pm\infty}\vec{H}(x,z)=H_{\mathrm{ext}}\vec{e}_z
\label{randb2}
\end{equation}
and 
\begin{equation}
  \begin{split}
    \left[(\vec{B}^{(3)}-\vec{B}^{(2)})
       \cdot \vec{n}^{(d)}\right]_{z=\zeta^{(d)}}& = 0\\[2ex]
    \left[(\vec{H}^{(3)}-\vec{H}^{(2)})
       \times \vec{n}^{(d)}\right]_{z=\zeta^{(d)}}& = 0
  \end{split}    
  \label{randb1a}
\end{equation}
\begin{equation}
  \begin{split}  
    \left[(\vec{B}^{(2)}-\vec{B}^{(1)})
        \cdot \vec{n}^{(0)}\right]_{z=\zeta^{(0)}}& = 0\\[2ex]
    \left[(\vec{H}^{(2)}-\vec{H}^{(1)})
        \times \vec{n}^{(0)}\right]_{z=\zeta^{(0)}}& = 0,
  \end{split}
  \label{randb1b}
\end{equation}
where $\vec{n}^{(0)}$ and $\vec{n}^{(d)}$ denote the normal vectors on
the lower and upper interface respectively. 
Note that the last four boundary conditions have to be fulfilled at
the free interfaces of the ferrofluid layer. They hence describe the
feedback of the interface modulations on the magnetic field. 
Note also that therefore the energy (\ref{energief}) depends in a
complicated non-local way on the surface deflections 
$\zetau(x)$ and $\zetao(x)$.

It is useful to introduce for each of the three liquid layers a scalar
magnetic potential $\phiu$, $\phim$, and $\phio$ respectively. The
potentials are related to the corresponding magnetic fields by 
\begin{equation}
  \Ha^{(i)}=-\nabla \Phi^{(i)}
\end{equation}
and as a consequence of (\ref{maxwell}) they obey the Laplace
equations 
\begin{equation}\label{laplace}
  \Delta \Phi^{(i)}=0.
\end{equation}
The boundary conditions (\ref{randb1a}) and (\ref{randb1b}) for $\Ha$
and $\B$ translate in the well-known way into conditions for the
continuity of the potentials themselves and jumps of their normal
derivatives \cite{LL}. 

It is furthermore convenient to measure all distances in units of the
inverse critical wavenumber  
\begin{equation}
  \frac{1}{k_{c,R}}=\sqrt{\frac{\sigmao}{\left(\rhom-\rhoo\right) g}}
\end{equation}
of the Rosensweig instability on an infinitely deep ferrofluid layer, all 
magnetic fields in units of the corresponding critical Rosensweig field 
\begin{equation}
  H_{c,R}=\sqrt{\frac{\left(1+\chi\right)\left(2+\chi\right)
      2\sqrt{\left(\rhom-\rhoo\right) g \sigmao}}{\chi^2\mu_0}},
  \label{hk}
\end{equation}
and energies per area in units of $\sigmao$. Moreover we introduce the 
parameter ratios
\begin{equation}
  \begin{split}
    \rho_1=\frac{\rhou}{\rhom},\quad\rho_3=\frac{\rhoo}{\rhom}\\[2ex]
    \sigma=\frac{\sigmau}{\sigmao},\quad\eta=\frac{\chi}{\chi+2}
  \end{split}
  \label{abk}
\end{equation}
with $\eta$ now characterizing the magnetic properties of the
ferrofluid. After rescaling the magnetic potentials according to 
\begin{align}
 -\frac{\left(2+\chi\right)}{\chi H_{\mathrm{ext}}}\Phi^{(1)}
      &\rightarrow \Phi^{(1)}\\[1ex]
 -\frac{\left(1+\chi\right)\left(2+\chi\right)}
               {\chi H_{\mathrm{ext}}}\Phi^{(2)}
      &\rightarrow \Phi^{(2)}\\[1ex]
 -\frac{\left(2+\chi\right)}{\chi H_{\mathrm{ext}}}\Phi^{(3)}
      &\rightarrow \Phi^{(3)}
 \label{phi}
\end{align}
the energy (\ref{energief}) assumes the dimensionless form 
\begin{equation}
  \begin{split}
    E\!\left[\zetau\!(x),\zetao\!(x)\right] &=\\[1ex]
    &\hspace{-16ex}\left\langle\frac{1}{2}\left[
      \left(\frac{\rho_1-1}{1-\rho_3}\right)\zeta^{(0)^2}
        + \zeta^{(d)^2}\right] \right.\\[1ex]
    &\hspace{-14ex}\left.  -H_{\mathrm{ext}}^2\left(\Phi^{(2)}
         \big|_{z=\zeta^{(d)}(x)} - 
        \Phi^{(2)}\big|_{z=\zeta^{(0)}(x)}\right)\right.\\[1ex]
    &\hspace{-14ex}\left. + \sigma\sqrt{1+(\partial_x\zeta^{(0)}(x))^2} 
    + \sqrt{1+(\partial_x\zeta^{(d)}(x))^2} \right\rangle.
  \end{split} 
  \label{energief1}  
\end{equation}
The boundary conditions (\ref{randb1a}) and (\ref{randb1b}) translate 
into 
\begin{equation}\label{bc1}
  \begin{split}
    \left[\partial_x\!(\Phi^{(3)}-\Phi^{(2)})
       \partial_x\zeta^{(d)} -
       \partial_z\!(\Phi^{(3)}-\Phi^{(2)}) 
     \right]_{z=\zeta^{(d)}} &= 0\\[3ex]
    \left[\frac{1+\eta}{1-\eta}\phio-\phim\right]_{z=\zetao} &= 0  
  \end{split}
\end{equation}
and 
\begin{equation}\label{bc2}
  \begin{split}
    \left[\partial_x\!(\phim-\phiu)
       \partial_x\zetau - \partial_z\!(\phim-\phiu)
     \right]_{z=\zetau} &= 0 \\[3ex]
    \left[\phim-\frac{1+\eta}{1-\eta}\phiu\right]_{z=\zetau} &= 0
  \end{split}
\end{equation}
respectively whereas the asymptotic boundary conditions (\ref{randb2})
acquire the form 
\begin{equation}
  \label{condinfty}
  \lim_{z\rightarrow +\infty}\partial_z\Phi^{(3)}\!
       \left(x,z\right)=
  \lim_{z\rightarrow -\infty}\partial_z\Phi^{(1)}\!\left(x,z\right)
  =\frac{1}{\eta}\quad.
\end{equation}


\section{Linear Stability Analysis} 

In this section we study the linear stability of the reference state with 
flat interfaces $\zetau\equiv 0$ and $\zetao\equiv d$. To this end we use 
the ansatzes 
\begin{equation}\label{ansatze1}
 \begin{split}
  \zetau&=A_1 \cos(kx)\\[1ex]
  \zetao&=d + B_1 \cos(kx)
 \end{split}
\end{equation}
for the interface profiles. The corresponding forms of the magnetic
potentials are then in view of (\ref{laplace}) and (\ref{condinfty}) 
\begin{equation}
\begin{split}\label{ansatze2}
  \phio & =
  \frac{z}{\eta}-\frac{2d}{1+\eta}+u_1 \E^{-k z}\cos(k x)\\[1ex]
  \phim & =
  \frac{z}{\eta}+\left(v_1^{+}\E^{k z}+
    v_1^{-}\E^{-k z}\right)\cos(k x)\\[1ex]
  \phiu & =
  \frac{z}{\eta}+w_1 \E^{k z}\cos(k x)\quad . 
\end{split}
\end{equation}

By using the linearized version of the boundary conditions (\ref{bc1}) and 
(\ref{bc2}) we can express the amplitudes $u_1,\ v_1^{+},\ v_1^{-}$, 
and $w_1$ in terms of $A_1$ and $B_1$. This then allows us to expand 
the energy (\ref{energief1}) up to second order in $A_1$ and
$B_1$ with the result\\[1ex] \rule{0mm}{7mm}
\begin{widetext}
\begin{align}\nonumber
    E\!\left(A_1,B_1\right) = &E\left(0,0\right)+\frac{1}{4}\!
        \left(\frac {\rho_1-1}{1-\rho_3}-2\,H_{\mathrm{ext}}^2 k 
        \frac{\eta\, \E^{-2\,d k}-1}{\eta^2\,\E^{-2\,d k}-1}
      +\sigma\,k^2 \right)\! A_1^2 \\[1ex]
    &+ \frac{1}{4}\!\left(1 -2H_{\mathrm{ext}}^2 k \frac{\eta\, 
       \E^{-2\,d k}-1}{\eta^2\, \E^{-2\,d k}-1}+k^{2}\right)\!B_1^2
    + H_{\mathrm{ext}}^2 k \left(\eta-1\right)\frac{\E^{-d k}}{\eta^2\,
       \E^{-2\,d k}-1}\,A_1 B_1.
    \label{lineng}
\end{align}
The energy has clearly a stationary point at $A_1=B_1=0$. It is stable
as long as the Hessian 
  \begin{equation}\label{Hessian}
    {\cal H}= \left(
      \begin{matrix}
        \frac{1}{2}\!\left(\frac {\rho_1-1}{1-\rho_3}-2\,H_{\mathrm{ext}}^2 k 
        \frac{\eta\, \E^{-2\,d k}-1}{\eta^2\,\E^{-2\,d k}-1}
      +\sigma\,k^2\right) & H_{\mathrm{ext}}^2 k 
   \left(\eta-1\right)\frac{\E^{-d k}}{\eta^2\,\E^{-2\,d k}-1} \\[1.5ex]
     H_{\mathrm{ext}}^2 k \left(\eta-1\right)\frac{\E^{-d k}}{\eta^2\,
       \E^{-2\,d k}-1} & \frac{1}{2}\!
   \left(1 -2H_{\mathrm{ext}}^2 k \frac{\eta\, 
       \E^{-2\,d k}-1}{\eta^2\, \E^{-2\,d k}-1}+k^{2}\right)
      \end{matrix}
    \right)
\end{equation}
\end{widetext}
is positive definite. An instability is signaled by a vanishing 
determinant of ${\cal H}$. 

We note that, of course, exactly the same condition results from the
usual procedure  
of linear stability analysis. In this case one investigates the  
dispersion relation $\omega(k)$ of interface deformations of the form 
$\zetau=A_1 \exp(i(kx-\omega t))$ and 
$\zetao=d+B_1 \exp(i(kx-\omega t))$ resulting from a linearization  
of the equations of motion. An instability occurs if $\omega$ acquires a 
positive imaginary part. The linearized equation of motion corresponds 
to the quadratic approximation of the energy. An advantage of the 
energetic approach is that it applies equally well to inviscid and 
viscous fluids. On the other hand it does not give information on the 
linear growth rate of the unstable perturbation. 

\begin{figure}[t]
  \centering
  \includegraphics[scale=0.95]{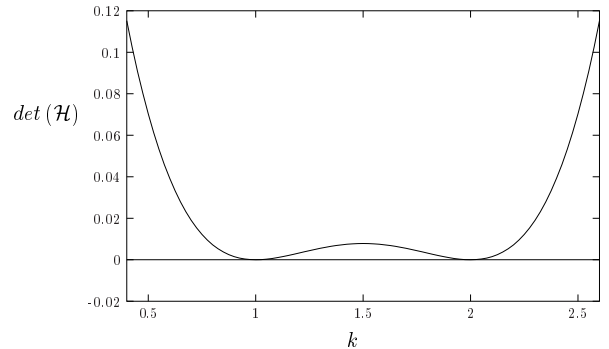}
  \caption{Determinant of the Hessian for an infinitely thick ferrofluid
    layer as function of the dimensionless wave number $k$. 
    Parameters $\rho_1=2$, $\rho_3=0.5$, $\sigma=0.5$, 
    $\eta=1.5$, are chosen such that the two independent
    interfaces get unstable at the same value of the magnetic field,
    (by definition $H_c=1$), but at different wave numbers 
    $k_c^{(u)} =1$ for the upper interface and $k_c^{(l)}=2$ for the
    lower one.}
  \label{fig:disprinfty}
\end{figure}

If the layer thickness $d$ tends to infinity it can be inferred from 
(\ref{Hessian}) that the off-diagonal elements of $\cH$ tend to zero 
whereas the diagonal elements reduce to the well-known form of a 
usual Rosensweig instability on a infinitely deep layer of ferrofluid 
\cite{Ros}. As expected we hence find in this limit two 
{\em uncoupled} interfaces showing {\em independent}  
Rosensweig instabilities of the usual kind. The situation is depicted 
in Fig.~\ref{fig:disprinfty} where we have shown the determinant of the 
Hessian matrix $\cH$ as function of the wavenumber $k$ for the case in
which the critical field of the two instabilities coincides but the
respective critical wavenumbers do not.

\begin{figure}[t]
  \centering
  \includegraphics[scale=0.95]{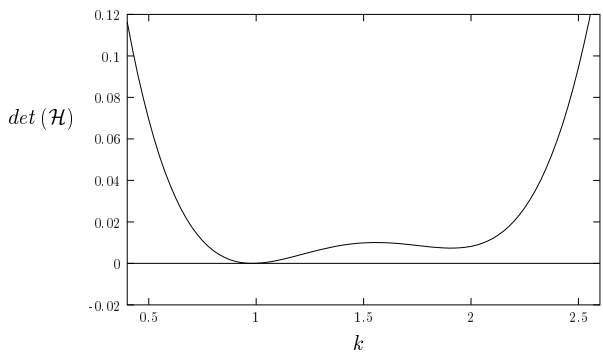}
  \caption{Same as Fig.~\ref{fig:disprinfty} for a layer thickness 
    $d=2$. The coupling between the two surfaces now lifts the 
    degeneracy characteristic of Fig.~\ref{fig:disprinfty} giving rise
    to new critical values for the wave vector, $k_c=0.96$, and the
    magnetic field, $H_c=0.98$.} 
  \label{dispersion}
\end{figure}

The situation changes if the layer thickness is reduced as shown in 
Fig.~\ref{dispersion}. Due to the interaction between the surface 
deformations mediated by the magnetic field the degeneracy observed in 
the case $d=\infty$ is lifted and the lower layer ``slaves'' the
upper one to its critical wavenumber. At the same time the critical
wavenumber is shifted somewhat, $k_c\neq 1$, from its ``pure''
value of the decoupled case. The same holds true for the critical
magnetic field strength. Moreover, the two interface deflections
accommodate to each other in an {\em anti-phase} fashion. This
manifests itself in different {\em signs} of $A_1$ and $B_1$ building
the components of the eigenvector corresponding to the zero eigenvalue
of ${\cal H}$. This anti-phase orientation was to be expected 
intuitively since it allows the largest gain in magnetic energy 
(cf. Fig.\ref{rosensweig}). 

\begin{figure}[htbp]
  \centering
  \includegraphics[scale=0.95]{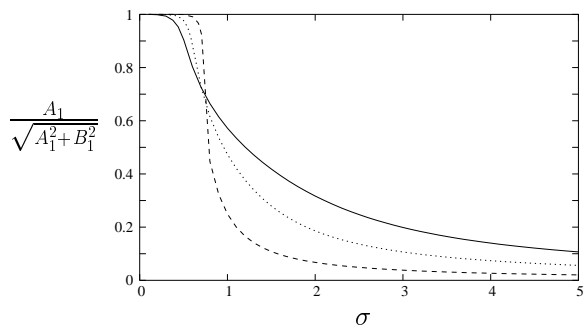}
  \caption{Relative amplitudes of the unstable modes as function of 
    the ratio $\sigma$ of the interface tensions as defined in 
    (\ref{abk}). The layer thickness is d=2 (dashed), d=1 (dotted), and 
    d=0.5 (solid). The other parameter values are $\eta=0.66$,
    $\rho_1=1.2$ and $\rho_3=0.85$.}
  \label{crossover}
\end{figure}

\begin{figure}[htbp]
  \centering
  \includegraphics[scale=0.95]{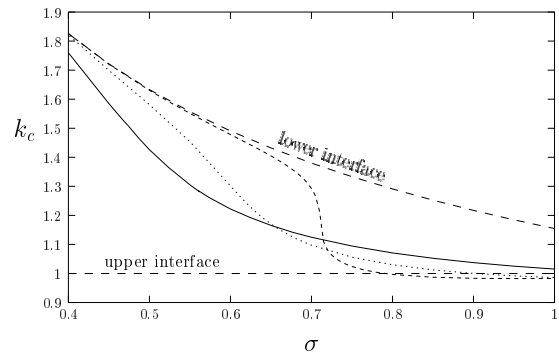}
  \caption{Dimensionless critical wavenumber of the linear instability as function 
   of the ratio $\sigma$ of interface tensions for layer thickness $d=\infty$ 
   (long dashed), $d=2$ (dashed), $d=1$ (dotted), $d=0.5$
   (solid). Other parameters as in fig.\ref{crossover}.}
  \label{fig:crossover-k}
\end{figure}

\begin{figure}[htbp]
  \centering
  \includegraphics[scale=0.95]{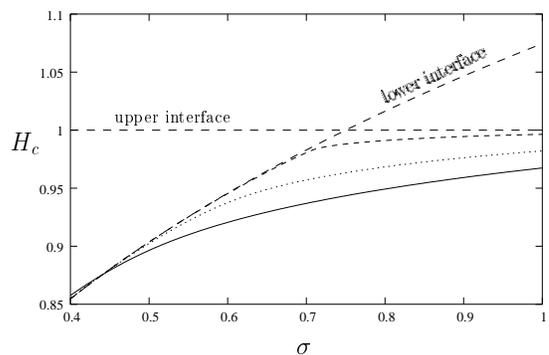}
  \caption{Dimensionless critical magnetic field $H_c$ of the linear instability as 
     function of the ratio $\sigma$ between the interface tensions for layer thickness 
     $d=\infty$ (long dashed), $d=2$ (dashed), $d=1$ (dotted), and 
    $d=0.5$ (solid). Other parameters as in fig.\ref{crossover}.}
  \label{fig:crossover-H}
\end{figure}

Which interface dominates which depends on the parameter values 
of the system and accordingly a crossover can be observed when 
some parameter is changed. In 
Figs.~\ref{crossover}-\ref{fig:crossover-H} we give some examples 
of such crossover phenomena when the ratio $\sigma$ between the two
interface tensions is changed. 
Fig.~\ref{crossover} displays the relative
amplitude of the two surface deflections. The figure 
clearly indicates that for small values of $\sigma$, i.e. when 
$\sigmau \ll \sigmao$, the lower instability dominates, $A_1 \gg B_1$, 
whereas with increasing $\sigma$ the amplitude of the lower interface 
deflection decreases and the coupled unstable modes get more and more 
dominated from the upper interface. 
Similarly Figs.~\ref{fig:crossover-k} and \ref{fig:crossover-H} show 
the crossover of the critical wavenumber $k_c$ and the critical field 
$H_c$, respectively, when $\sigma$ is varied. In all cases the crossover 
gets sharper with increasing depth $d$ of the ferrofluid layer as 
expected.

Although the linear analysis already reveals some aspects of the 
interplay between the two Rosensweig instabilities it is not able to
yield information about the static pattern  
of interface deflections that will eventually emerge. In order to address 
this problem we need to extend our analysis to include  
non-linear terms able to saturate the exponential growth predicted by 
the linear stability theory. This is the subject of the next section. 


\section{Weakly Nonlinear Analysis}

In our analysis of the energy of the surface deflections the
instability of the flat surface corresponds to the minimum of the
energy at $A_1=B_1=0$ turning into a saddle point. Within the
quadratic approximation the energy hence decreases down to $-\infty$
with increasing amplitudes $A_1$ and $B_1$. In reality, however,
already for moderate values of $A_1$ and $B_1$ higher order terms in the
expansion of the energy have to be included which cure this
divergence. As a result the energy again {\em in}creases with 
increasing amplitudes $A_1$ and $B_1$ and correspondingly a new
minimum forms describing the new stationary surface profiles
$\zetau(x)$ and $\zetao(x)$. 

We assume that the susceptibility of the ferrofluid is sufficiently
small such that
an expansion of the energy (\ref{energief1}) up to fourth order in the
amplitudes of the surface deflection is sufficient to find the new
stationary state. Such an expansion is equivalent to the derivation of
a third order amplitude equation for the unstable mode \cite{FrEn2}. 
In order to obtain a consistent expansion the Fourier expansions
(\ref{ansatze1}) and (\ref{ansatze2}) have to be extended according to 
\begin{equation}
\begin{split}  
  \zetao\left(x\right) & = d+ \sum_{n=1}^{2} B_n
    \cos\left(n  k x\right)\\[1ex] 
  \zetau\left(x\right) & = \sum_{n=1}^{2} A_n\cos\left(n  k x\right)
\end{split}
\label{ansatze3}
\end{equation}
and 
\begin{equation}
\begin{split}
  \phio & =\frac{z}{\eta}-\frac{2d}{1+\eta}+\sum_{n=1}^{2} u_n 
        \E^{-n  k z}\cos(n  k x)\\[1ex]
  \phim & = \frac{z}{\eta}+\sum_{n=1}^{2} \left(v_n^{+}\E^{n  k z}+
       v_n^{-}\E^{-n  k z}\right)\cos(n  k x)\\[1ex]
  \phiu & =  \frac{z}{\eta}+\sum_{n=1}^{2} w_n 
      \E^{n  k z}\cos(n  k x) .
\end{split}
\label{ansatze4}
\end{equation}

To explicitly perform the minimization of the free energy a variant of
the computer algebra code documented in \cite{Rene} is used. Fixing
the desired order of the expansion (four in our case) this program 
selects in a first step those amplitude combinations which are
compatible with the translational invariance of the problem in
$x$-direction. In our case only 17 of the originally 70 terms remain 
after this procedure. In a second
step the ansatzes (\ref{ansatze3}) and (\ref{ansatze4}) are used 
in the boundary conditions (\ref{bc1}) and (\ref{bc2}) and the
coefficients $u_n,v_n^+,v_n^-,w_n$ are determined as polynomials in the
$A_n$ and $B_n$. After this the energy (\ref{energief1}) can be
expanded up to fourth order in $A_1$ and $B_1$ and up to second order
in $A_2$ and $B_2$. Several of the remaining terms disappear after
the integration over $x$ implicit in the horizontal average in
(\ref{energief1}). Minimizing the resulting expression in
$A_2$ and $B_2$ we find that both are of order $A_1^2, B_1^2$ which
proofs the consistency of our expansion {\it a-posteriori}. Finally
the free energy is minimized in the amplitudes $A_1$ and $B_1$ of the
main modes. The final expressions are explicit but too long to be
displayed. 

For $d=\infty$ we again reproduce the results obtained for the
standard Rosensweig instability on a layer of infinite depth 
\cite{Gal,FrEn}. For $d<\infty$ the two interfaces couple and the two
surface deflections arrange in a stable, anti-phase pattern. 

\begin{table}\caption{\label{table1} Magnetic fluid parameters used in
  Figs. \ref{fig:amp1}-\ref{fig:amp3}}
  \begin{tabular}{|lcl|lcl|} \hline
    \multicolumn{3}{|c|}{Experimental parameters} &
    \multicolumn{3}{|c|}{Dimensionless values}\\ \hline\hline 
    $\rhou$&=&$1.69\; \mathrm{g/cm^3}$ & && \\ 
    $\rhom$&=& $1.12\; \mathrm{g/cm^3}$
    &\raisebox{1.5ex}[-1.5ex]{$\rho_1$}&\raisebox{1.5ex}[-1.5ex]{=}& 
    \raisebox{1.5ex}[-1.5ex]{$1.51$}\\ 
    $\rhoo$&=&$0.0013\; \mathrm{g/cm^3} $ &
    \raisebox{1.5ex}[-1.5ex]{$\rho_3$}&\raisebox{1.5ex}[-1.5ex]{=}& 
    \raisebox{1.5ex}[-1.5ex]{$0.001$} \\ 
    $\sigmau$&=&$16.6\; \mathrm{mN/m} $ &&& \\
    $\sigmao$&=& $25.9\; \mathrm{mN/m}$ &
    \raisebox{1.5ex}[-1.5ex]{$\sigma$}&\raisebox{1.5ex}[-1.5ex]{=}& 
    \raisebox{1.5ex}[-1.5ex]{$0.64$}\\
    $d$&=&$1.54\; \mathrm{mm}$ & $d$&=&$1.0 $\\
    $\chi$&=&$0.8 $ & $\eta$&=&$0.29 $\\ \hline
  \end{tabular}
\end{table}

\begin{figure}[htbp]
  \centering
  \includegraphics[scale=0.95]{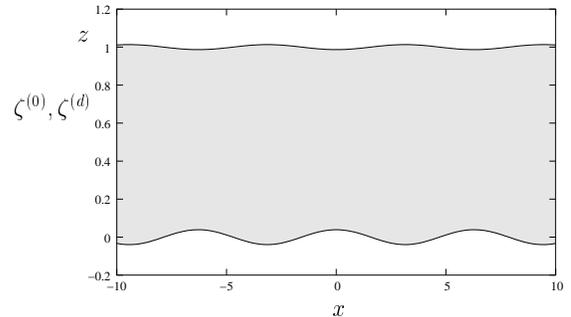}
  \caption{Stationary pattern of coupled surface deflections that
    evolve after the instability of the state with flat interfaces. 
    The ferrofluid layer is shown in gray. 
    Parameters are given in table \ref{table1}, the value of the
    external magnetic field is $H_{\mathrm{ext}}=1.0001  H_c$. The figure 
    uses dimensionless units.}
  \label{fig:amp1}
\end{figure}

\begin{figure}[htbp]
  \centering
  \includegraphics[scale=0.95]{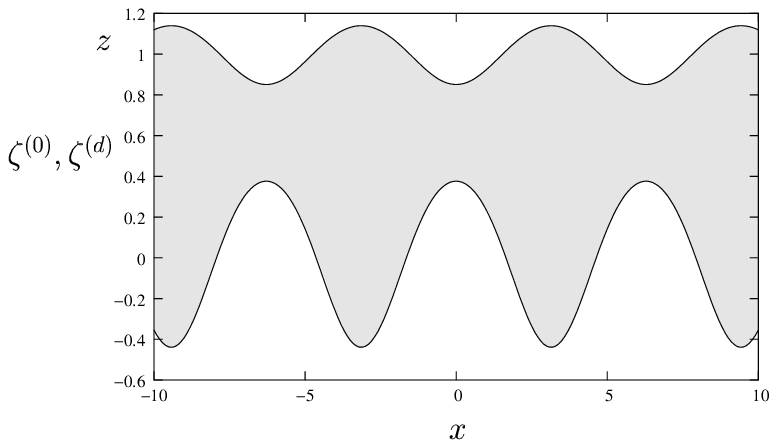}
  \caption{Same as fig.\ref{fig:amp1} for $H_{\mathrm{ext}}=1.01 H_c$.} 
  \label{fig:amp2}
\end{figure}\begin{figure}[htbp]
  \centering
  \includegraphics[scale=0.95]{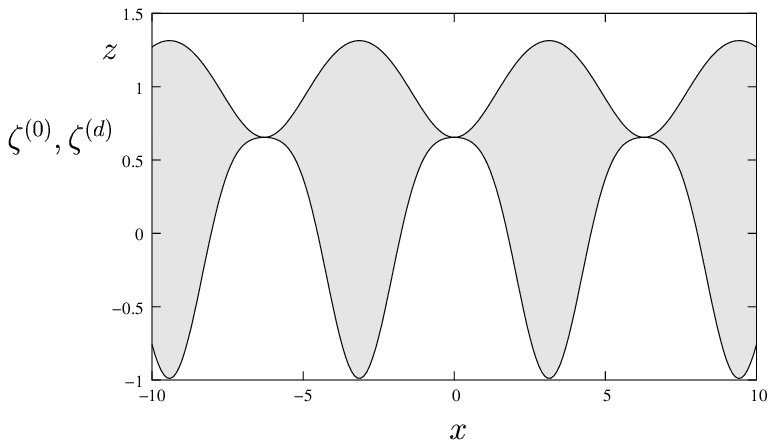}
  \caption{Same as fig.\ref{fig:amp1} for  $H_{\mathrm{ext}}=1.04 H_c$.} 
  \label{fig:amp3}
\end{figure}

To elucidate this final structure in detail we consider the case of
experimentally realistic parameters collected in table \ref{table1}. 
From the linear stability analysis we find for the dimensionless wave
number $k_c=0.84$ and for the corresponding dimensionless field 
$H_c=0.75$. The stationary interface profiles resulting from the
weakly non-linear analysis are 
displayed in Figs.~\ref{fig:amp1}-\ref{fig:amp3}. As can be seen the
lower interface is the dominating one. For slightly overcritical
magnetic field the lower interface already shows an array of developed
Rosensweig ridges whereas the upper one is just gently curved by the
inhomogeneous magnetic pressure resulting from the field modulation
induced by the lower interface (Fig.~\ref{fig:amp1}). With increasing
field both deformations grow (Fig.~\ref{fig:amp2}). A particular
interesting case is shown in Fig.~\ref{fig:amp3} where the
amplitudes of the surface deflections have increased to such an extent
that the two interfaces touch each other. Correspondingly
the ferrofluid layer stays no longer connected but
disintegrates. In our two-dimensional ($x,z$) model this gives rise to
the formation of parallel slices. In a more realistic
three-dimensional setting, including surface variations in
$y$-direction as well, the layer would evolve into a regular array of
disconnected {\em islands}. A similar phenomenon occurs in the usual
Rosensweig instability on very shallow layers of ferrofluid \cite{FrEn}. 

\begin{figure}[htbp]
  \centering
  \includegraphics[scale=1]{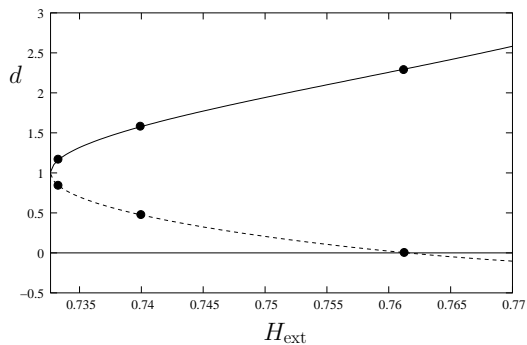}
  \caption{Maximum (full line) and minimum (dashed line) dimensionless thickness of
    the ferrofluid layer as function of the dimensionless external field
    strength for the parameters given in table \ref{table1}. The dots
    correspond to the situations displayed in 
    Figs.~\ref{fig:amp1}-\ref{fig:amp3} respectively. For the field at
    which the minimum distance between the interfaces shrinks to zero
    the layer disintegrates into an array of disconnected rolls.}
    \label{fig:bifurcation}
\end{figure}

In Fig.~\ref{fig:bifurcation} we have shown the maximum and the
minimum layer thickness as function of the magnetic field. The
formation of islands occurs when the lower branch intersects with the
horizontal axis. Note that, at least for the parameters of table
\ref{table1}, this happens already for a field exceeding the critical
one by only 4\%.

With the appearance of such an island structure our theoretical model
breaks down. To study the future evolution of the structure when
increasing the field still further it is more appropriate to start
from a model of independent ferrofluid drops \cite{BPS}.

\begin{figure}[htbp]
  \centering
  \includegraphics[scale=0.95]{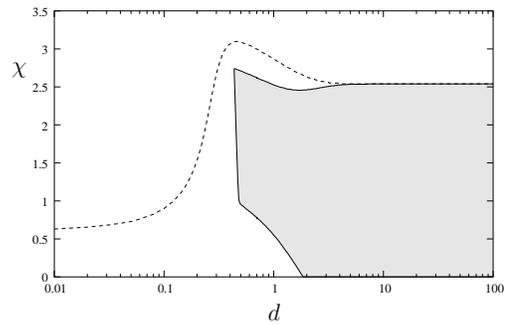}
  \caption{Region of consistency of our nonlinear treatment of the
    pattern formation in the plane spannend by the dimensionless layer
    thickness $d$ and the susceptibility $\chi$. Outside the shaded
    region the fourth order terms in the expansion of the energy are
    not sufficient to saturate the linear instability and higher order
    terms are needed to get finite results for the amplitudes of
    interface deflections when minimizing the energy. The dashed line
    is the result of \cite{FrEn} for a ferrofluid layer with rigid
    bottom. Parameter values are from table \ref{table1}.}
  \label{mud}
\end{figure}

In omitting higher orders of the expansion of the energy in the
amplitudes of surface deflection in our nonlinear analysis we have
tacitly assumed that the fourth order terms are sufficient to saturate the
linear instability, i.e. to make $E[\zetau, \zetao]\to\infty$ for 
$((\zetau)^2+(\zetao)^2) \to \infty$. This, however, is correct only if the 
susceptibility $\chi$ of the ferrofluid is not too large and the
thickness $d$ of the layer is not too small. In Fig.~\ref{mud} we have displayed the
region in the $d$-$\chi$ plane, in which our treatment is consistent. 

\begin{figure}[htbp]
  \centering
  \includegraphics[scale=1]{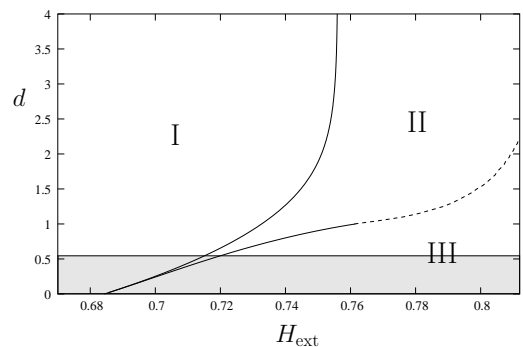}
  \caption{Phase diagram in the plane spanned by the dimensionless external field
    $H_{\rm ext}$ and layer thickness $d$ for a ferrofluid sandwich structure with the
    parameters given in table \ref{table1}. In region I both
    interfaces are flat, in region II, ridges occur and in region III,
    the layer disintegrates into an array  of disconnected rolls. 
    The dashed line indicates that for larger  values of the magnetic
    field higher order terms are necessary to accurately determine the
    location of the transition line. In the shaded region the
    fourth order terms in the energy are not sufficient to saturate
    the linear instability and higher order terms are mandatory, cf. 
    Fig.~\ref{mud}. } 
  \label{fig:inseln}
\end{figure}

Finally Fig.~\ref{fig:inseln} gives the phase diagram of the
ferrofluid sandwich structure showing the transition lines from 
flat interfaces to anti-phase interface modulations and further to
disconnected regions. It would be interesting to compare the location
of these theoretical lines with experimental results. 

\section{Conclusion}

In the present paper we have investigated the linear and weakly
nonlinear theory of two coupled Rosensweig instabilities in a
ferrofluid sandwich structure. To this end an approximate expression
for the energy of the system was minimized in the deflection
amplitudes of the two interfaces between magnetic and non-magnetic
liquids. The approximate expression for the free energy was obtained
from a fourth order perturbative expansion in these interface
deflections. At the onset of instability the two individual Rosensweig
instabilities compete and depending on the concrete values of the
parameters one is able to ``slave'' the other one to its unstable
wavenumber. As a result, a stable antiphase pattern of two interacting
modulated interfaces arises. For sufficiently thin layers and
sufficiently large magnetic fields the two curved interfaces may touch
each other which brings about the disintegration of the layer and
gives rise to disconnected rolls or islands. Using realistic parameter
values we gave estimates of the required layer
thicknesses and magnetic fields necessary to observe this phenomenon
in an experiment. Being perturbative in nature our theoretical
analysis has a limited range of validity which we quantified by
estimating the contributions of higher order terms. It is possible
though tedious to push the expansion to higher orders in a systematic
way.

\begin{acknowledgments}
We would like thank Reinhard Richter and Norbert Buske for arousing
our interest in ferrofluid sandwich structures and Ren\'e Friedrichs
for numerous helpful discussions. This work was supported by {\it Deutsche
  Forschungsgemeinschaft} under grant FOR/301.
\end{acknowledgments}

\end{document}